# Tuning the competition between superconductivity and charge order in kagome superconductor Cs(V$_{1-x}$Nb$_x$)$_3$Sb$_5$

Yongkai Li[1,3,4,†], Qing Li[2,†], Xinwei Fan[2,†], Jinjin Liu[1,3], Qi Feng[1,3], Min Liu[1,3], Chunlei Wang[5], Jia-Xin Yin[6], Junxi Duan[1,3], Xiang Li[1,3,4], Zhiwei Wang[1,3,4,*], Hai-Hu Wen[2,*], Yugui Yao[1,3,*]

[1]Centre for Quantum Physics, Key Laboratory of Advanced Optoelectronic Quantum Architecture and Measurement (MOE), School of Physics, Beijing Institute of Technology, Beijing 100081, P. R. China.

[2]National Laboratory of Solid State Microstructures and Department of Physics, Collaborative Innovation Center of Advanced Microstructures, Nanjing University, Nanjing 210093, P. R. China

[3]Beijing Key Lab of Nanophotonics and Ultrafine Optoelectronic Systems, Beijing Institute of Technology, Beijing 100081, P. R. China.

[4]Material Science Center, Yangtze Delta Region Academy of Beijing Institute of Technology, Jiaxing, 314011, P. R. China

[5]Key Laboratory of Microelectronics and Energy of Henan Province, School of Physics and Electronic Engineering, Xinyang Normal University, Xinyang 464000, P. R. China

[6]Laboratory for Topological Quantum Matter and Advanced Spectroscopy (B7), Department of Physics, Princeton University, Princeton, New Jersey, USA.

[†]These authors contributed equally to this work.

[*]e-mail: zhiweiwang@bit.edu.cn; hhwen@nju.edu.cn; ygyao@bit.edu.cn

The recently discovered coexistence of superconductivity and charge density wave order in the kagome systems $A$V$_3$Sb$_5$ ($A$ = K, Rb, Cs) has stimulated enormous interest. According to theory, a vanadium-based kagome system may host a flat band, nontrivial linear dispersive Dirac surface states and electronic **correlation. Despite intensive investigations, it remains controversial about the origin of the charge density wave (CDW) order, how does the superconductivity relate to the CDW, and whether the anomalous Hall effect (AHE) arises primarily from the kagome lattice or the CDW order. We report an extensive investigation on $Cs(V_{1-x}Nb_x)_3Sb_5$ samples with systematic Nb doping. Our results show that the Nb doping induces apparent suppression of CDW order and promotes superconductivity; meanwhile, the AHE and magnetoresistance (MR) will be significantly weakened together with the CDW order. Combining with our density functional calculations, we interpret these effects by an antiphase shift of the Fermi energy with respect to the saddle points near M and the Fermi surface centered around $\Gamma$. It is found that the former depletes the filled states for the CDW instability and worsens the nesting condition for CDW order; while the latter lifts the Fermi level upward and enlarges the Fermi surface surrounding the $\Gamma$ point, and thus promotes superconductivity. Our results uncover a delicate but unusual competition between the CDW order and superconductivity.**

**Introduction**

The recently discovered kagome superconductor $A$V$_3$Sb$_5$ ($A$ = K, Rb, Cs) family has attracted tremendous attention due to its novel properties, including possible chiral charge density wave (CDW), anomalous Hall effect (AHE), and unconventional superconductivity (SC), etc [1-11]. Resistivity, magnetic susceptibility, and specific heat show an anomaly [1, 12] at about 90K for CsV$_3$Sb$_5$. A gapped feature can be observed at the similar temperature by the angle-resolved photoemission spectroscopy (ARPES)[13-15], tunneling spectrum[16, 17], far-infrared optics[18]; thus all these experiments reveal that a CDW transition occurs at the temperature T$_{CDW}$. Because there are nesting wave-vectors between different saddle points near M points, it was argued that the CDW might be induced by this saddle point nesting effect. This can get partial support from ARPES measurements with the gapped feature being maximized near M points[13, 14]. The resistive and Hall effect measurements under a magnetic field illustrate a coincident temperature of both the CDW effect and the AHE [1, 12], manifesting a similar origin. The Fourier transform intensity on the topography measured by a scanning tunneling microscope (STM) under magnetic fields with opposite vorticities shows a rotational symmetry breaking, which was explained as the existence of a chiral current state with a time-reversal symmetry breaking (TRSB) [19]. However, this effect was not observed in the repeating experiments of other groups [20]. A clear in-plane twofold symmetry of the normal state electronic properties has been reported by the c-axis resistivity measurements, which shows an intact relation with the CDW order [21, 22]. Concerning superconductivity, multiple gaps with probably nodeless structure[17, 23] have been found with the maximum ratio of $2\Delta / k_B T_c \approx 5.5$, indicating a relatively strong coupling for the superconductivity. Despite these extensive investigations, it remains controversial whether and how the superconductivity is related to the CDW order, and whether is the CDW order intrinsically correlated with the AHE or not.

In this paper, we report the successful growth of high-quality Nb-doped Cs(V$_{1-x}$Nb$_x$)$_3$Sb$_5$ (CVNS) single crystals, and found that CDW is suppressed, but superconductivity is enhanced. With the increase of Nb doping level, CDW ordering phenomena and its temperature are suppressed. Meanwhile, $T_c$ is enhanced monotonically to about 4.45K, which is the highest $T_c$ at ambient pressure in the bulk crystal of the $A$V$_3$Sb$_5$ family so far. At the same time, AHE and MR are significantly weakened together with the weakening of the CDW order, indicating that the AHE has an intimate relationship with the CDW and the dominated electric conduction changes from multiband to single-band induced by Nb doping. Combining with the density functional theory (DFT) calculations, we find that the doping will induce an antiphase shift of the Fermi energy with respect to the saddle points near M and the electron Fermi pocket surrounding Γ. Thus, we propose that the influence of doping on superconductivity and the CDW order does not resemble the usual picture of competition concerning the effective density of states (DOS). Our results indicate that the dominant pairing instability may arise from the large Γ-centered electron Fermi surface, while the CDW order results from the nesting effect of the saddle points near the Fermi energy.

Nb-doped Cs(V$_{1-x}$Nb$_x$)$_3$Sb$_5$ single crystals were synthesized with different doping level ranging from $x$ = 0 to 0.07. We first characterized the crystal structure of CVNS single crystals by XRD measurements. Fig. S1(a) in the Supplementary Information (SI) shows a schematic illustration of crystal structure for Nb-doped CsV$_3$Sb$_5$ after structure refinement. Fig. S1(b) presents a typical x-ray diffraction (XRD) pattern of the crystal with $x$ = 0.07, and only (00$l$) diffraction peaks with a very narrow full-width at half-maximum (FWHM) are seen, which confirms a pure phase of the as-grown crystals. The typical lateral size of the crystals is about 4×4 mm$^2$, with a clear hexagonal morphology, as shown in the inset of Fig. S1(b). The XRD patterns for all samples are quite similar, as shown in Fig. S2. To check whether Nb is doped to the V-site or not after doping, the crystal with $x$ = 0.07 was used for structure resolution and refinement by using the SHELXTL software package after taking a single-crystal XRD measurement. We found Nb was successfully doped into V-site as expected, and the crystal structure is presented in Fig. S1(a) with the partial occupation of the V sites with Nb. The detailed structural information is summarized in Table S1 and S2 in the SI. The lattice parameters $a$ (=$b$) and $c$ are refined to be 5.5157 Å and 9.3083 Å, respectively.

Compared to those of parent CsV$_3$Sb$_5$ ($a$ = 5.4949 Å and $c$ = 9.3085 Å) [1], we can see that the lattice parameter $a$ becomes slightly larger while $c$ keeps almost the same versus doping. This means Nb doping tends to affect the lattice constant in the $ab$ plane, rather than in the $c$-axis. This is why (00$l$) peaks did not show obvious shift with Nb doping (see Fig. S2). From the energy-dispersive-spectroscopy (EDS) pattern, as illustrated in Fig. S1(c), we can see that there is a peak arising from Nb, which verifies that Nb does have been doped into CsV$_3$Sb$_5$. The Nb peak appears in all doped samples, while the parent sample does not have that peak, as shown in Fig. S3. The observed suppression of CDW and enhancement of T$_c$, which will be discussed in the following section, further confirm that Nb is successfully doped into the lattice effectively. We would like to point out that the doping amount $x$ adopted in the whole text reflects the actual Nb content, which was estimated from the EDS analysis, and we found that the solubility limit of Nb in CVNS is about 7%.

Now let us look at how SC and CDW orders evolve with Nb doping. Fig 1 shows an apparent competition between SC and CDW at various Nb doping levels. It is obvious that with the increase of Nb content, not only the CDW transition temperature decreases from 92 K for $x$ = 0 sample to 58 K for $x$ = 0.07 sample, but also the amplitude of d$\rho$/d$T$, which indicates the CDW transition, becomes weaker and weaker, as shown in Fig. 1(a). However, even in the sample with the highest Nb doping content (i.e., $x$ = 0.07), the CDW order was not completely suppressed. Meanwhile, SC transition temperature was enhanced gradually and reached a maximum $T_c$ of about 4.45 K in $x$ = 0.07 sample (see Fig. 1a), which is the highest $T_c$ in the bulk crystal of $A$V$_3$Sb$_5$ family at ambient pressure so far. Here the $T_c$ is chosen as the temperature corresponding to the midpoint of the resistivity drop. Therefore, the CDW and SC compete with each other in the Nb-doped CNVS system. This competitive relationship was also observed in pressurized $A$V$_3$Sb$_5$ [8-10]. The residual resistivity ratio (RRR) for $x$ = 0 samples is 83, which indicates a good quality of this crystal. The RRR becomes smaller and smaller when more Nb are doped into CsV$_3$Sb$_5$; this is most likely because the scattering effect becomes stronger caused by doping. Fig 1(c) shows the temperature dependence of magnetic susceptibility near the SC transition. Both zero-field-cooled (ZFC) and field-cooled (FC) measurements were performed under an applied external magnetic field of 5 Oe for all CVNS samples. A diamagnetic signal can be seen for all measured samples. The SC transition temperature increases with the increase of Nb doping content, which is consistent with that in resistivity measurements. In most undoped samples, the superconducting transition measured by magnetic susceptibility looks not very sharp. Still, in the present Nb-doped samples, this transition looks quite sharp, and a complete magnetic shielding appears at low temperatures, suggesting good quality of the samples.

Fig 1(d) summarizes the phase diagram of CVNS single crystals, where the effect of Nb doping on CDW and SC is presented. One can see that $T_{CDW}$ is suppressed with increasing Nb doping while $T_c$ is enhanced simultaneously, but they are coexisting in all doped samples, which features competing and coexisting behavior. This is a bit different from those in hole-doped $CsV_{3-x}Ti_xSb_5$ and $CsV_3Sb_{5-x}Sn_x$, where the CDW was wholly suppressed at a higher doping levels [24, 25]. Furthermore, $T_c$ increases monotonically in CVNS with increasing doping while there are two SC regions in both $CsV_{3-x}Ti_xSb_5$ and $CsV_3Sb_{5-x}Sn_x$ systems[24, 25]. However, there is also a group that reported only one SC dome was observed in $CsV_{3-x}Ti_xSb_5$ with comparable Ti doping level [26].

To check how the upper critical field $\mu_0H_{c2}$ evolves in Nb-doped CVNS, we performed the temperature dependent resistivity measurements under various applied external magnetic fields for $x = 0$ and 0.07 crystals with the fields perpendicular to the *ab* plane, as presented in Fig. 2(a) and 2(b). For both crystals, the $T_c$ is gradually suppressed with an increased magnetic field. For the parent $CsV_3Sb_5$, superconductivity was almost entirely killed at 1.8 K for $\mu_0H = 0.5$ T. And for the sample with $x = 0.07$, to demonstrate the $\mu_0H_{c2}$ better, the sample was cooled down to 0.03 K in a dilution fridge. The zero resistivity can still be achieved at 0.03 K for $\mu_0H = 3$ T. The data points in Fig. 2(c) is taken from the midpoint of each resistivity drop in Fig. 2(a) and 2(b). The two-band theory fits the two sets of data very well and gives the $\mu_0H_{c2}(0)$ to be about 1.1 T for $x = 0$ sample and 3.2 T for $x = 0.07$ sample, which corresponds to the coherence length $\xi = \sqrt{\Phi_0/2\pi\mu_0 H_{c2}}$ to be 17.2 nm and 10.1 nm, respectively, where $\Phi_0$ is the magnetic flux quantum.

AHE was observed commonly in all $A$V$_3$Sb$_5$ family compounds [2, 4, 11], and it was explained that the chiral flux phase might exhibit a 2×2 charge order, which breaks time-reversal symmetry and results in AHE state [3]. However, the experimental evidence for this scenario is only reported in the parent CsV$_3$Sb$_5$ at high pressures [5]. Our Hall resistivity measurements on CVNS provide robust evidence for this scenario at ambient pressure. Fig. 3(a) presents the magnetic field dependence of Hall resistivity at various temperatures for the parent CsV$_3$Sb$_5$ sample. At high temperatures above 50 K, $\rho_{xy}$ exhibits a linear behavior, indicating that the scattering is dominated by a single band. And with the temperature decreasing from 300 K, the slop of $\rho_{xy}$ ($\mu_0 H$) becomes larger and larger until 90 K. Then it gets smaller and smaller; below 50K, $\rho_{xy}$ gradually deviates from the linear relationship with the temperature decreasing and changes its sign around 30K. Such a phenomenon could be attributed to the significant enhancement of hole mobility at low temperature due to the multiband feature of CsV$_3$Sb$_5$. An antisymmetric sideways "S" line shape is observed in the low field region, which is similar to that reported by Yu *et al*. in CsV$_3$Sb$_5$ [12] and by Yang *et al*. in the K$_{1-x}$V$_3$Sb$_5$ [2].

Fig. 3(b) and 3(c) show Hall resistivity at various temperatures for Cs(V$_{1-x}$Nb$_x$)$_3$Sb$_5$ samples with $x$ = 0.02 and 0.07. It is clear that with Nb doping, the antisymmetric sideways "S" line shape in the low field region will fade out, indicating that the AHE is getting weaker and weaker together with the CDW order. This phenomenon provides direct evidence that the AHE has an intimate relationship with the CDW. Another conspicuous phenomenon is that the slope of $\rho_{xy}$ ($\mu_0 H$) in the sample with $x$ = 0.07 is always negative, and no sign change was observed through the whole temperature range. This indicates that the conduction changes from a multiband dominated feature to a single-band dominated one in Nb-doped system, and the dominated carriers turn to be electron like. The Hall coefficient ($R_H$) and carrier density

($n_e$) are shown in Fig. S4, which coincides with what we discussed above. The magnetoresistance (MR) becomes smaller gradually with the increasing of Nb doping, as shown in Fig. S5. This is also understandable since the magnetoresistance will become weaker when the electric conduction of the system is dominated by a single band after Nb doping, being consistent well with the Hall signal discussed above. Fig. 3(d) summarizes the extracted anomalous Hall signal $\rho_{yx}^{AHE}$ by subtracting the local linear ordinary Hall background at 5 K, which clearly shows that the AHE was significantly weakened by Nb doping. This shows that the anomalous Hall effect has an intimate relationship with the CDW.

We have shown the resistive and magnetic data measured on the Nb-doped CVNS samples. It is clear that, with the doping of Nb, the CDW effect and its temperature are significantly weakened. Meanwhile, the Hall signal changes from a multiband dominated transport to a single band dominated one. This can also get corroboration from the magnetoresistance data. To have a comprehensive understanding of these systematic evolutions, we have taken a calculation based on the density functional theory by using the lattice constants obtained on our samples. The key results are shown in Fig. 4. We can see a straightforward modification of the band structure after Nb doping, as shown in Fig. 4(a) by the red dotted line compared to the undoped band structure shown in the blue solid line. Two prominent features can be observed here as the effect of Nb doping. Firstly, the major momentum area due to the van Hove singularity (VHS) near the M point, namely the saddle point is slightly below the Fermi energy in the undoped samples; by doping Nb to the system, however, this VHS band moves up and crosses the Fermi energy, as shown in Fig. 4(b). This behavior significantly modifies the Fermi surface giving rise to a broken hole on the Fermi surface in the path between Γ and M point as shown in Fig. 4(c), since the saddle point band no longer crosses the Fermi level along this path after doping. Similar behavior was also observed in Ti-doped samples and was considered to be responsible for the absence of the CDW order [26]. Thus, the doping depletes the filled density of states of the VHS near M, which weakens the instability of the saddle points leading to suppression of the CDW effect and the CDW temperature. Secondly, the Fermi energy is lifted a little bit near Γ, which enlarges the electron like Fermi surface centered around Γ. However, in contrast to a rigid shift of the Fermi energy to the band structure via hole doping by the element of the same period of vanadium like Ti, doping with the element of the same group in the periodic table, like Nb, may not add to the system more electrons. Rather it gives a more complex modification to the band structure, namely an antiphase phase shift of the Fermi energy near M and Γ points. We would like to point out that this antiphase shift was observed in the recent ARPES measurements by Kato *et al* [27]. In this sense, we would conclude that the competition between the CDW and superconductivity in the Nb-doped samples does not adopt the usual picture that they compete for effective DOS across the Fermi surface. This occurs via a delicate but unusual competition through modifying the bands near Γ and M points.

The picture proposed here can also be corroborated by the Nb-doping induced evolution of the Hall effect and magnetoresistance. In the undoped phase, we have several bands crossing the Fermi energy, and this leads to a multiband effect, which can be detected by the temperature-dependent Hall coefficient and its sign change below a certain temperature [1, 12]. Meanwhile, the magnetoresistance in the undoped sample is huge due to the involvement of the different charge carrier densities, scattering rate, and mobilities, like that in the clean $MgB_2$ system[28]. By doping Nb to the system, the bands corresponding to the VSH near the M points rise above the Fermi energy, leading to a reduced contribution to the electric conduction. Meanwhile, the electron-like Fermi surface centered around Γ is getting enlarged, which enhances the contribution, and gradually becomes dominant in the electric conduction. The CDW in the undoped phase is established through the instability of the nesting effect of the momentum space near the saddle points around M. Concerning the anomalous Hall effect, our results clearly show that it is intimately related to the CDW phase. While it remains to be resolved what is the final reason for the AHE. It may arise from the orbital loop current phase as predicted by some theories[3], or it is induced by the field-induced correlated magnetic moments of the vanadium ions[29], which deserves further study.

In summary, we have successfully synthesized Nb-doped $Cs(V_{1-x}Nb_x)_3Sb_5$ single crystals with a solution limit of 7%. We found that the Nb-doping gradually suppresses CDW order and enhances SC with the highest $T_c$ of 4.45 K, which indicates the competition of CDW and SC. At the same time, AHE and MR were significantly weakened together with CDW caused by Nb-doping, which provides experimental evidence for the close relationship between CDW and AHE. This competition between CDW and SC can be interpreted nicely by our DFT calculations on the delicate modifications to the bands centered around $\Gamma$ and the VHS bands near M points. It is found that the Nb-doping drives the VHS bands from below Fermi energy to above, which depletes the filled DOS near M points and weakens the instability for the CDW phase; meanwhile, the electron-like Fermi surface centered around $\Gamma$ is enlarged and wins more DOS for the superconducting pairing. Our combined experimental and theoretical results reveal a delicate but unusual competition of the CDW and superconducting phase.

**Acknowledgements**

Z. Wang acknowledges helpful discussions with Yaomin Dai, Takafumi Sato, and Xun Shi. The calculations were performed on the High-Performance-Computing (HPC) clusters of Collaborative Innovation Center of Advanced Microstructures in Nanjing University. This work was supported by the National Key R&D Program of China (2020YFA0308800), National Natural Science Foundation of China (No. 92065109, 11534005, 11927809, NSFC-DFG12061131001), Beijing Natural Science Foundation (Nos. Z210006 and Z190006), and Beijing Institute of Technology Research Fund Program for Young Scholars (No. 3180012222011). Z.W. thanks the Analysis & Testing Center at BIT for assistance in facility support.

# References

[1]  B. R. Ortiz, L. C. Gomes, J. R. Morey, M. Winiarski, M. Bordelon, J. S. Mangum, I. W. H. Oswald, J. A. Rodriguez-Rivera, J. R. Neilson, S. D. Wilson, E. Ertekin, T. M. McQueen, and E. S. Toberer, New kagome prototype materials: discovery of $KV_3Sb_5$, $RbV_3Sb_5$, and $CsV_3Sb_5$, Phys. Rev. Mater. **3**, 094407 (2019).

[2]  S.-Y. Yang, Y. Wang, B. R. Ortiz, D. Liu, J. Gayles, E. Derunova, R. Gonzalez-Hernandez, L. Šmejkal, Y. Chen, S. S. P. Parkin, S. D. Wilson, E. S. Toberer, T. McQueen, and M. N. Ali, Giant, unconventional anomalous Hall effect in the metallic frustrated magnet candidate, $KV_3Sb_5$, Sci. Adv. 6, eabb6003 (2020).

[3]  X. Feng, K. Jiang, Z. Wang, and J. Hu, Chiral flux phase in the Kagome superconductor $AV_3Sb_5$, Sci. Bull. **66**, 1384–1388 (2021)

[4]  Q. Yin, Z. Tu, C. Gong, Y. Fu, S. Yan, and H. Lei, Superconductivity and Normal-State Properties of Kagome Metal $RbV_3Sb_5$ Single Crystals, Chin. Phys. Lett. **38**, 037403 (2021).

[5]  F. H. Yu, X. K. Wen, Z. G. Gui, T. Wu, Z. Wang, Z.J. Xiang, J. Ying, and X. Chen, Pressure tuning of the anomalous Hall effect in the kagome superconductor $CsV_3Sb_5$, Chin. Phys. B **31** 017405 (2022)

[6]  H. Chen, H. Yang, B. Hu, Z. Zhao, J. Yuan, Y. Xing, G. Qian, Z. Huang, G. Li, Y. Ye, S. Ma, S. Ni, H. Zhang, Q. Yin, C. Gong, Z. Tu, H. Lei, H. Tan, S. Zhou, C. Shen, X. Dong, B. Yan, Z. Wang, and H. J. Gao, Roton pair density wave in a strong-coupling kagome superconductor, Nature **599**, 222-228 (2021).

[7]  Z. Wang, Y. X. Jiang, J. X. Yin, Y. Li, G. Y. Wang, H. L. Huang, S. Shao, J. Liu, P. Zhu, N. Shumiya, M. S. Hossain, H. Liu, Y. Shi, J. Duan, X. Li, G. Chang, P. Dai, Z. Ye, G. Xu, Y. Wang, H. Zheng, J. Jia, M. Z. Hasan, and Y. Yao, Electronic nature of chiral charge order in the kagome superconductor $CsV_3Sb_5$, Phys. Rev. B **104**, 075148 (2021).

[8]  K. Y. Chen, N. N. Wang, Q. W. Yin, Y. H. Gu, K. Jiang, Z. J. Tu, C. S. Gong, Y. Uwatoko, J. P. Sun, H. C. Lei, J. P. Hu, and J. G. Cheng, Double Superconducting Dome and Triple Enhancement of $T_c$ in the Kagome Superconductor $CsV_3Sb_5$ under High Pressure, Phys. Rev. Lett. **126**, 247001 (2021).

[9]  F. H. Yu, D. H. Ma, W. Z. Zhuo, S. Q. Liu, X. K. Wen, B. Lei, J. J. Ying, and X. H. Chen, Unusual competition of superconductivity and charge-density-wave state in a compressed topological kagome metal, Nat. Commun. **12**, 3645 (2021).

[10] Q. Wang, P. Kong, W. Shi, C. Pei, C. Wen, L. Gao, Y. Zhao, Q. Yin, Y. Wu, G. Li, H. Lei, J. Li, Y. Chen, S. Yan, and Y. Qi, Charge Density Wave Orders and Enhanced Superconductivity under Pressure in the Kagome Metal $CsV_3Sb_5$, Adv. Mater. **33**, 2102813 (2021).

[11] B. R. Ortiz, S. M. L. Teicher, Y. Hu, J. L. Zuo, P. M. Sarte, E. C. Schueller, A. M. M. Abeykoon, M. J. Krogstad, S. Rosenkranz, R. Osborn, R. Seshadri, L. Balents, J. He, and S. D. Wilson, $CsV_3Sb_5$: A $Z_2$ Topological Kagome Metal with a Superconducting Ground State, Phys. Rev. Lett. **125,** 247002 (2020).

[12] F. H. Yu, T. Wu, Z. Y. Wang, B. Lei, W. Z. Zhuo, J. J. Ying, and X. H. Chen, Concurrence of anomalous Hall effect and charge density wave in a superconducting topological kagome metal, Phys. Rev. B **104**, L041103 (2021).

[13] Z. Wang, S. Ma, Y. Zhang, H. Yang, Z. Zhao, Y. Ou, Y. Zhu, S. Ni, Z. Lu, H. Chen, K. Jiang, L. Yu, Y. Zhang, X. Dong, J. Hu, H. J. Gao, and Z. Zhao, Distinctive momentum dependent charge-

density-wave gap observed in $CsV_3Sb_5$ superconductor with topological Kagome lattice, arXiv:2104.05556

[14] K. Nakayama, Y. Li, T. Kato, M. Liu, Z. Wang, T. Takahashi, Y. Yao and T. Sato, Multiple energy scales, and anisotropic energy gap in the charge-density-wave phase of the kagome superconductor $CsV_3Sb_5$, Phys. Rev. B **104**, L161112 (2021).

[15] K. Nakayama, Y. Li, T. Kato, M. Liu, Z. Wang, T. Takahashi, Y. Yao, and T. Sato, Carrier Injection and Manipulation of Charge-Density Wave in Kagome Superconductor $CsV_3Sb_5$, Phys. Rev. X **12**, 011001 (2022)

[16] Z. Liang, X. Hou, F. Zhang, W. Ma, P. Wu, Z. Zhang, F. Yu, J. J. Ying, K. Jiang, L. Shan, Z. Wang, and X. H. Chen, Three-Dimensional Charge Density Wave and Surface-Dependent Vortex-Core States in a Kagome Superconductor $CsV_3Sb_5$, Phys. Rev. X **11**, 031026 (2021).

[17] H. S. Xu, Y. J. Yan, R. Yin, W. Xia, S. Fang, Z. Chen, Y. Li, W. Yang, Y. Guo, and D. L. Feng, Multiband superconductivity with sign-preserving order parameter in kagome superconductor $CsV_3Sb_5$, Phys. Rev. Lett. **127**, 187004 (2021).

[18] X. Zhou, Y. Li, X. Fan, J. Hao, Y. Dai, Z. Wang, Y. Yao, and H. Wen, Origin of charge density wave in the kagome metal $CsV_3Sb_5$ as revealed by optical spectroscopy, Phys. Rev. B **104**, L041101 (2021).

[19] Y. X. Jiang, J. X. Yin, M. M. Denner, N. Shumiya, B. R. Ortiz, G, Xu, Z. Guguchia, J. He, M. S. Hossain, X. Liu, J. Ruff, L. Kautzsch, S. S. Zhang, G. Chang, I. Belopolski, Q. Zhang, T. A. Cochran, D. Multer, M. Litskevich, Z. J. Cheng, X. P. Yang, Z. Wang, R. Thomale, T. Neupert, S. D. Wilson, and M. Z. Hasan, Unconventional chiral charge order in kagome superconductor $KV_3Sb_5$, Nat. Mater. **20**, 1353-1357 (2021).

[20] H. Li, S. Wan, H. Li, Q. Li, Q. Gu, H. Yang, Y. Li, Z. Wang, Y. Yao, and H. H. Wen, No observation of chiral flux current in the topological kagome metal $CsV_3Sb_5$, Phys. Rev. B **105**, 045102 (2022).

[21] Y. Xiang, Q. Li, Y. Li, W. Xie, H. Yang, Z. Wang, Y. Yao, and H. H. Wen, Twofold symmetry of c-axis resistivity in topological kagome superconductor $CsV_3Sb_5$ with in-plane rotating magnetic field, Nat. Commun. **12**, 6727 (2021).

[22] M. M. Denner, R. Thomale, and T. Neupert, Analysis of charge order in the kagome metal $AV_3Sb_5$ (A = K, Rb, Cs), Phys. Rev. Lett. **127**, 217601 (2021).

[23] W. Duan, Z. Nie, S. Luo, F. Yu, B. R. Ortiz, L. Yin, H. Su, F. Du, A. Wang, Y. Chen, X. Lu, J. Ying, S. D. Wilson, X. Chen, Y. Song, and H. Yuan, Nodeless superconductivity in the kagome metal $CsV_3Sb_5$, Sci. China Phys. Mech. Astron. **64**, 107462 (2021).

[24] Y. M. Oey, B. R. Ortiz, F. Kaboudvand, J. Frassineti, E. Garcia, S. Sanna, V. F. Mitrović, R. Seshadri, and S. D. Wilson, Fermi level tuning and double-dome superconductivity in the kagome metals $CsV_3Sb_{5-x}Sn_x$, Preprint at https://arxiv.org/abs/2110.10912 (2021).

[25] H. Yang, Y, Zhang, Z. Huang, Z. Zhao, J. Shi, G. Qian, B. Hu, Z. Lu, H. Zhang, C. Shen, X. Lin, Z. Wang, S. J. Pennycook H. Chen, X. Dong, W. Zhou, and H. J. Gao, Doping and two distinct phases in strong-coupling kagome superconductors, Preprint at https://arxiv.org/abs/2110.11228 (2021).

[26] Y. Liu, Y. Wang, Y. Cai, Z. Hao, X. M. Ma, L. Wang, C. Liu, J. Chen, L. Zhou, J. Wang, S. Wang, H. He, Y. Liu, S. Cui, J. Wang, B. Huang, C. Chen, and J. W. Mei, Doping evolution of superconductivity, charge order and band topology in hole-doped topological kagome superconductors $Cs(V_{1-x}Ti_x)_3Sb_5$, Preprint at https://arxiv.org/abs/2110.12651 (2021).

[27]  T. Kato, Y. Li, K. Nakayama, Z. Wang, S. Souma, F. Matsui, M. Kitamura, K. Horiba, H. Kumigashira, T. Takahashi, Y. Yao, and T. Sato, Fermiology and Origin of Tc Enhancement in a Kagome Superconductor Cs(V$_{1-x}$Nb$_x$)$_3$Sb$_5$. （will be submitted）

[28]  H. Yang, Y. Liu, C. Zhuang, J. Shi, Y. Yao, S. Massidda, M. Monni, Y. Jia, X. Xi, Q. Li, Z. K. Liu, Q. Feng, and H. H. Wen, Fully band-resolved scattering rate in MgB$_2$ revealed by the nonlinear hall effect and magnetoresistance measurements, Phys. Rev. Lett. **101**, 067001 (2008).

[29]  L. Balents, Spin liquids in frustrated magnets, Nature **464**, 199-208 (2010)

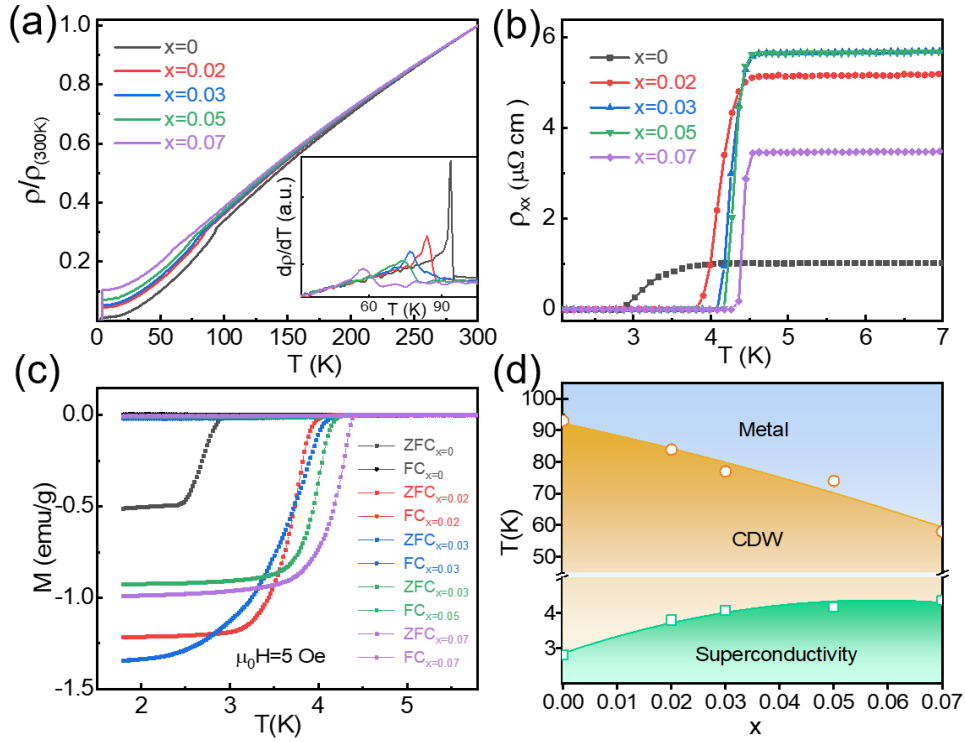

FIG. 1. CDW and SC in Cs(V$_{1-x}$Nb$_x$)$_3$Sb$_5$. (a), Temperature dependence of in-plane resistivity measured from 300 K to 1.8 K for CVNS. The data were normalized to the resistivity at 300K for respective samples; the inset shows d$\rho$/d$T$ as a function of temperature near CDW transition. It is clear that the CDW transition temperature gradually decreases with the increase of Nb doping. (b), $\rho_{xx}(T)$ curves at temperatures near the SC transition for CVNS, it is shown that T$_c$ increases monotonically with doping Nb. (c), Temperature dependence of magnetic susceptibility for CVNS measured with the applied field of 5 Oe, both ZFC and FC curves are presented. The variation trend of $T_c$ with Nb content is consistent with that in resistivity measurements. (d), Phase diagram of CVNS, which illustrates the competition between CDW and SC.

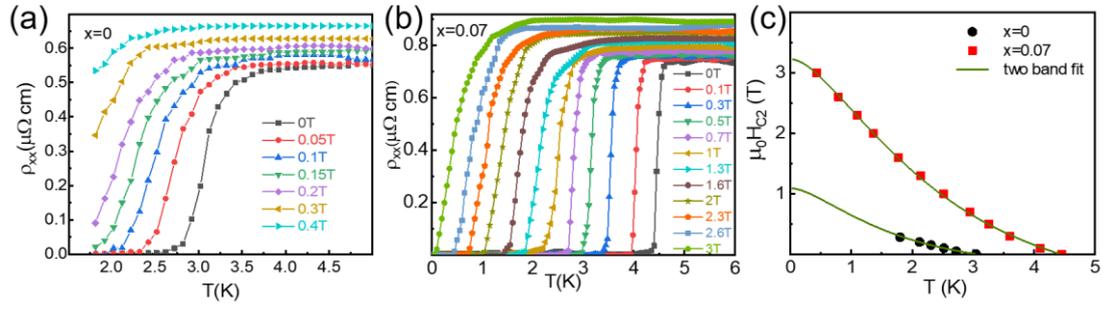

FIG. 2. Determination of $\mu_0H_{c2}$ in $Cs(V_{1-x}Nb_x)_3Sb_5$. (a), Temperature dependence of resistivity measured under various magnetic fields applied perpendicular to *ab* plane for parent $CsV_3Sb_5$ sample. (b), The same as in (a), but measured down to 30 mK for $x$ = 0.07 sample. (c), $\mu_0H_{c2}$ vs. T phase diagram determined from 50% $\rho_N$ (resistivity at normal state) in (a) and (b), the solid line shows the two-band fitting.

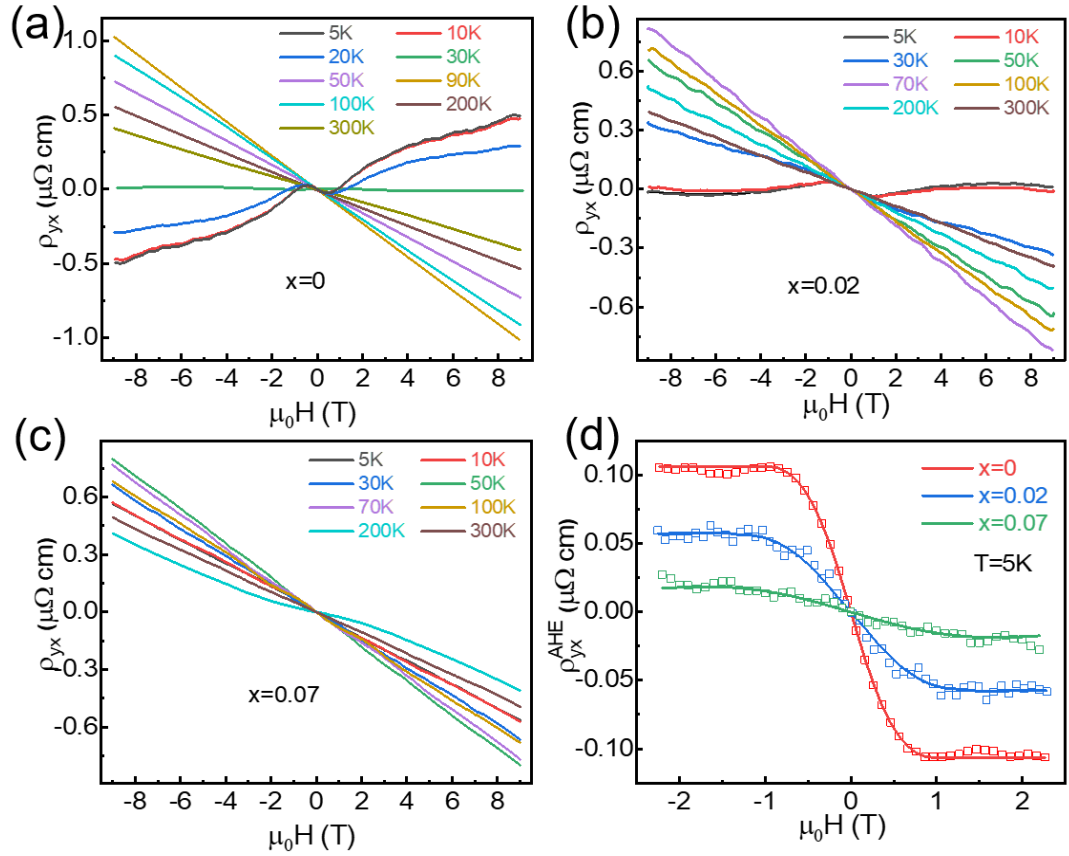

FIG. 3. Hall effect in $Cs(V_{1-x}Nb_x)_3Sb$. (a)-(c), The Hall resistivity measured with the current applied in the *ab* plane, and the magnetic field applied perpendicular to the *ab* plane at various temperatures for samples with *x* = 0, 0.02, and 0.07, respectively. The AHE shows up an antisymmetric "S" shape in the low-field region at low temperatures for *x* = 0 and 0.02 samples, but for *x* =0.07 sample, this "S" shape is almost disappeared even at 5 K. (d), Extracted $\rho_{yx}^{AHE}$ taken by subtracting the local linear ordinary Hall background at 5K. AHE was significantly weakened with the Nb doping.

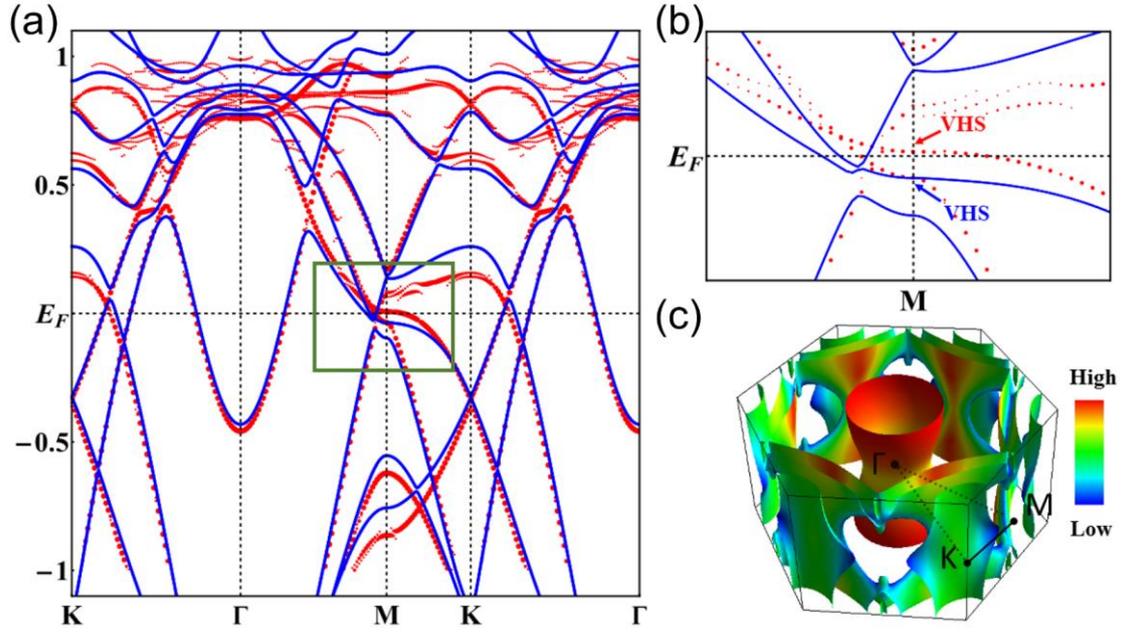

FIG. 4. Electronic structures of Cs(V$_{1-x}$Nb$_x$)$_3$Sb$_5$ ($x$ = 0.07). (a), The band structure of undoped CsV$_3$Sb$_5$ is shown with blue solid lines. The red dotted lines represent the unfolded band structure of the doped sample. The green frame indicates the magnified region of the band structure shown in (b). (c), The phenomenological Fermi surface is produced by shifting the Fermi energy in the undoped sample downwards about 44 meV. The color indicates the Fermi velocity. A broken hole appears on the Fermi surface near M points where the saddle points locate.